# Numerical simulation of baffled Circulating Fluidized Bed with Geldart B particles

Salma Benzarti[a*], Hatem Mhiri[a], Hervé Bournot[b]

[a]*Unit of Thermic and Thermodynamics of the Industrial Processes, department of energy, University of Monastir, Monastir, Tunisia*
[b],*Aix-Marseille Université, CNRS, IUSTI Marseille, France*
* E-mail address: salma_benzarti@yahoo.fr
  Tel: +21629344333

*Abstract*

   As most classical fluidization technologies, Circulating Fluidized Beds (CFB) are widely employed in chemical and physical operations. However, CFB Reactors still have some disadvantages such as the non-uniform solid particles distribution and the backflow of the solid particles near the wall, which deteriorate the system mixing, and lead to low system conversion/heat and mass transfer rate. To improve the CFB performance, we propose in the current work to study the effect of the ring baffle configuration on the fluidization system hydrodynamics using a multifluid Eulerian CFD model, incorporating the Kinetic Theory of Granular Flow. In this approach, two different types of ring baffles, circular and trapezoidal, were considered and the corresponding results were compared. The results revealed that the incorporation of ring baffles improved the system mixing and reduced the backflow near the wall. Nevertheless, the shape of baffles had a limited impact on the system hydrodynamics.

*Graphical abstract*

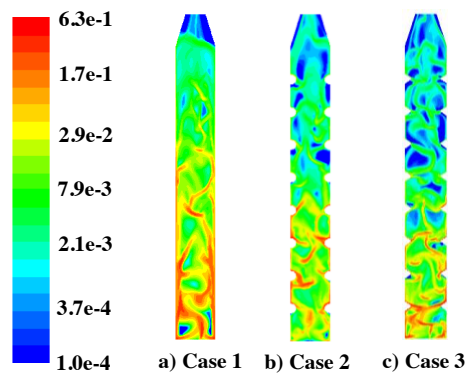

Instantaneous solid volume fraction in CFB riser: (a) without baffles (b) with circular ring baffles and (c) with trapezoidal ring baffles

*Highlights*

- The circulating fluidized bed with ring baffles was investigated.
- Eulerian–Eulerian approach with Gidaspow gas-solid drag model and standard k-e model was performed
- Two ring baffle configurations were investigated.
- Time-averaged results relative to circulating fluidized beds in presence and in absence of ring baffles were compared.

- The incorporation of ring baffles improved the system mixing and reduced the backflow near the wall

## Nomenclature

| | | |
|---|---|---|
| $C_{2\varepsilon}, C_{1\varepsilon}$ | Turbulence model Coefficients | - |
| $C_D$ | Drag coefficient | - |
| $D$ | Riser width | m |
| $d_s$ | Particle diameter | m |
| $e_s$ | Particle-particle restitution coefficient | - |
| $g$ | Gravitational acceleration, 9.81 | m s$^{-2}$ |
| $g_0$ | Radial distribution function | - |
| $G_{k,g}$ | Production of turbulent kinetic energy | W m$^{-3}$ |
| $G_s$ | Solids external mass flux | kg m$^{-2}$s$^{-1}$ |
| $H$ | Riser height | m |
| $I$ | Unit tensor | - |
| $I_{2D}$ | The second invariant of the deviatoric stress tensor | - |
| $J$ | Granular energy transfer | kg m$^{-1}$ s$^{-3}$ |
| $k_g$ | Turbulence quantities of gas phase | m$^2$ s$^{-2}$ |
| $P$ | Pressure | N m$^{-2}$ |
| $P_s$ | Solids pressure | N m$^{-2}$ |
| $q$ | Diffusion of fluctuating energy | kg s$^{-3}$ |
| $Re$ | Reynols number | - |
| $t$ | Time | s |
| $u$ | Velocity | m s$^{-1}$ |
| $u_{dr}$ | Drift velocity | m s$^{-1}$ |
| $U_g$ | Superficial gas velocity | m s$^{-1}$ |
| $q$ | Diffusion of fluctuating energy | kg s$^{-3}$ |

*Subscripts*

| | |
|---|---|
| KTGF | Kinetic Theory of Granular Flow |
| Col | Collisional |
| Kin | Kinetic |
| max | Maximum |
| fr | Frictional |
| X | Lateral coordinate, m |
| s | Solid phase |

g       Gas phase

*Greek Letters*

| | | |
|---|---|---|
| $\varepsilon_g$ | Turbulence dissipation of gas phase | m² s⁻³ |
| α | Volume fraction | - |
| ρ | Density | kg m⁻³ |
| $\xi_s$ | Solid bulk viscosity | Pa·s |
| φ | Specularity coefficient | - |
| $\sigma_k, \sigma_\varepsilon$ | Turbulent Prandtl number | - |
| β | Gas-particle interaction coefficient | Kg m⁻³ s⁻¹ |
| Θ | Granular temperature | m² s⁻² |
| τ | Stress tensor | N m⁻² |
| ϕ | Angle of internal friction | ° |
| γ | Collisional dissipation of energy | Kg m⁻³ s⁻¹ |

## 1. Introduction

Circulating Fluidized Beds (CFBs) are a well-known device that has several significant advantages due to their effective contacting reactors' property and the continuous process they provide, coupled with a high throughput. They consequently have been widely used in industrial processes, such as chemical, petrochemical, pharmaceutical, metallurgical and energy-related industries [1]. In spite of their widespread applications, CFBs still have some disadvantages, namely the non-uniform solid particles' distribution in both the radial and axial system directions as well as the back-fall of solid particles near the wall region, which deteriorate the system mixing and lead to low system conversion/heat and mass transfer rate [2,3].

One way to overcome the disadvantages of CFBs reactors is to install internal structures [4-8]. The main purpose of this solution is to scrape the down flowing solids away from the wall towards the center of the riser, leading to more uniform radial solid particles' distribution and intensify the gas-solids contact. There are several kinds of internal structures used in practice, such as ring baffles, perforated plates, tubes, inserted bodies, etc. Among all types of internal structures evaluated, ring baffles proved to highly enhance the radial mixing and reduce the back-mixing phenomenon with no effect from the overall system pressure drop. Jiang et al. [4], investigated the effect of ring baffles on the performance of a catalytic Circulating Fluidized Bed Reactor (CFBR). They observed that the inclusion of ring baffles could improve the performance of a CFBR for the ozone decomposition reaction. Their experimental results showed that in risers with ring baffles, the ozone concentration in the radial direction is much more uniform and the ozone conversion much higher at the lowest superficial gas velocity. Zhu et al. [6], investigated experimentally the effect of internals ring types with various open area ratios on axial pressure distribution and gas-solids flow structure in a 7.6 cm-diameter and 3 m- height CFBR riser. They found that the ring baffles could reduce radial non-uniformity, evenly reorder the solid particles' radial distribution and formed a denser region above in contrast with a more dilute region below the ring baffle. Therdthianwong et al. [7], conducted numerical studies in order to examine the hydrodynamics and ozone decomposition in CFBR with and without ring baffles. They found that the addition of baffles improve radial gas and solid particles' mixing and increase the ozone decomposition by approximately 5 to 12%. Samruamphianskun et al. [8], used a factorial design analysis to investigate the effect of ring baffles design on the hydrodynamics of a CFBR riser. They found that the ring baffle opening area, the inter-baffle space and the interaction between these two terms have a significant effect on the solid particles' distribution and system mixing. More recently, Samruamphianskun et al. [9] investigated the effect of the operating parameters on the system hydrodynamics and mixing inside two CFBR risers with different ring baffle configurations using Computational Fluid Dynamics (CFD) simulations and a factorial experimental design analysis. The results showed that the operating parameters that had a significant effect on the distribution of solid particles were the inlet gas velocity and solid particle mass flux. They also found that the use of baffles in the mixing zone approximates the flow of the stuck profile, breaking the annulus and redirecting the catalyst particles to the center of the riser. As a result a high conversion and a significant reduction of the over cracking of the products was observed on the walls. The insertion of baffles in the feed region brings then greater uniformity in the distribution of particles in this region and decreases the speed of the core region, thus reducing the product sub-conversion.

As a continuity to above mentioned studies, we propose to explore in the present work the effect of the ring baffle configuration on the fluidization system hydrodynamics in a CFB filled with Geldart B particles using a multifluid Eulerian CFD model, incorporating the Kinetic Theory of Granular Flow. In this context, two different types of ring baffles, circular and trapezoidal, are considered and the corresponding results are compared.

## 2. Computational Fluid Dynamics simulation set up

### 2.1. System description

Our simulations are based upon the experimental work of Zaabout et al. [10]. The case system, as illustrated schematically in Fig. 1(a), is an ordinary CFB riser without ring baffles, with a cross-section of $0.2 \times 0.2$ m$^2$ and a height of 2 m. The gas consists of air at ambient conditions while the solid particles are made of glass beads characterized by a density of 2400 kg/m3 and a sauter mean diameter of 120 μm (Geldart's Group B solid particles [11]). The static bed height is 0.1 m and the system is simulated at a cold flow condition without any chemical reaction. The inlet gas are uniformly fed to the system at the bottom of the CFB riser with a superficial gas velocity of 1 m/s. The solid particles are fed from the one side inlet with a width of 0.05 m. the inlet solid particles' velocity was calculated from the solid mass flux of a 0.22 kg/m$^2$s. The gas and solid particles exited the system from the top of the riser with a width of 0.1 m. The numerical model was validated in our previous work [12]. In order to investigate the effect of the ring baffle configuration on the system hydrodynamics, two types of ring baffle configurations (circular and trapezoidal) were tested. The two modified riser geometries were displayed as a simplified schematic diagram in Fig. 1(b) and (c), respectively. There are 8 pieces of ring baffles in each configuration with an open area of 70%. The different models are performed using the commercial computer aided design (CAD) program GAMBIT 2.2.30 and then exported to the commercial CFD simulation program ANSYS FLUENT 6.3.26. The computational domain of the risers used in the current work is discretized using a uniform quadratic mesh with 30864, 42284 and 41808 number of cells respectively for case 1, case 2 and case 3. The simulations are conducted unsteadily with a time step of 10$^{-4}$ s during 30 s.

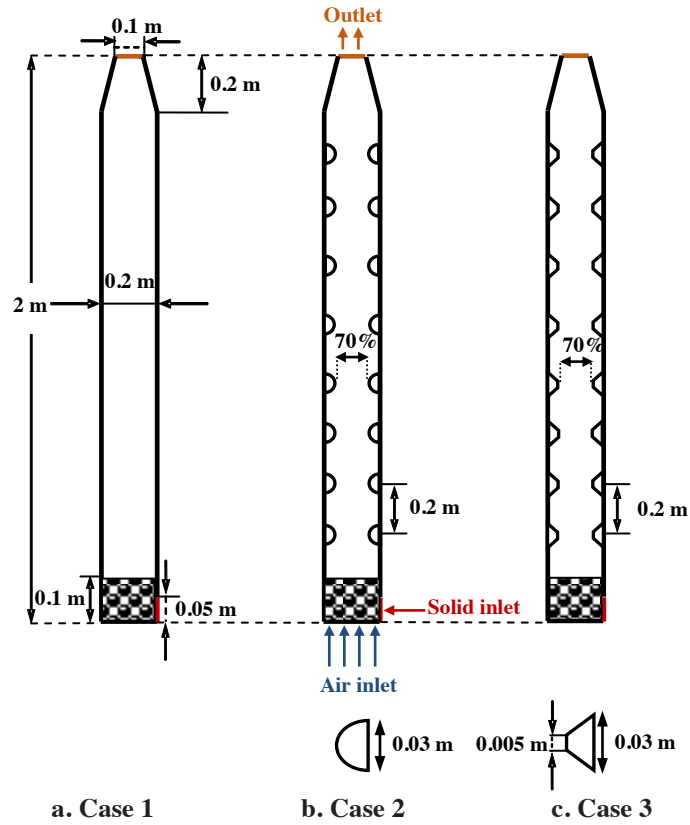

***Fig. 1.*** *Schema of the different CFBR tested geometries: (a) without baffles (b) with circular ring baffles and (c) with trapezoidal ring baffles*

### 2.2. Hydrodynamic model

To describe properly the 2D gas-solid flow, in the present study we propose to solve the governing equations of mass, momentum and granular energy for both the gas and solid phases by means of a transient Eulerian- Eulerian approach. In order to completely describe the governing equations, appropriate constitutive equation where specified to describe the physical properties in each phases. In the current work, we propose to use the equations based on the Kinetic Theory of Granular Flow (KTGF) developed by Gidaspow [13] and available in the commercial software package Ansys Fluent. This theory's validity has been proven by many researchers [14-18] and was founded on the application used of the kinetic theory of dense gases to the particulate assemblies. This approach gives more perspicacity in terms of interactions between particles by assuming that binary collisions between hard spheres take place instantaneously. To describe the solid fluctuating kinetic energy, this model introduces one conservation equation called the granular temperature. The physical properties of the solid phase can therefore be obtained as a function of the restitution coefficient and the granular temperature. The momentum exchange between the two phases is accounted for by the drag coefficient, which has a significant effect on the prediction of the Eulerian-Eulerian model. In this study, the drag model of Gidaspow [13] is used for this system, which has been shown to yield good predictions according to our previous work [12]. The dispersed turbulent model is applied to describe the effects of the turbulent fluctuations of the velocities and the scalar quantities in the gas phase. This model is based upon the standard k-epsilon turbulence model with supplementary terms for the momentum exchange between the two phases. The calculation of the turbulence for the dispersed phase is achieved by application of the Tchen theory correlation [19] into Fluent. These models have already been validated by comparison with the CFB experimental data of [10], and a perfect agreement between the computational and experimental data was already obtained [12]. The governing equations with their constitutive equations are provided in Table 1.

**Table 1**
A summary of the governing equations and constitutive equations

**A. Governing equations**

(a) Mass conservation equations of gas and solids phases

$$\frac{\partial(\alpha_g \rho_g)}{\partial t} + \nabla \cdot (\alpha_g \rho_g \vec{u}_g) = 0 \qquad (1)$$

$$\frac{\partial(\alpha_s \rho_s)}{\partial t} + \nabla \cdot (\alpha_s \rho_s \vec{u}_s) = 0 \qquad (2)$$

$$\alpha_g + \alpha_s = 1 \qquad (3)$$

(b) Momentum conservation equations of gas and solids phases

$$\frac{\partial(\alpha_g \rho_g \vec{u}_g)}{\partial t} + \nabla \cdot (\alpha_g \rho_g \vec{u}_g \vec{u}_g) = \nabla \cdot (\bar{\bar{\tau}}_g) - \alpha_g \nabla P - \beta(\vec{u_g} - \vec{u_s}) + \alpha_g \rho_g g \qquad (4)$$

$$\frac{\partial(\alpha_s \rho_s \vec{u}_s)}{\partial t} + \nabla \cdot (\alpha_s \rho_s \vec{u}_s \vec{u}_s) = \nabla \cdot (\bar{\bar{\tau}}_s) - \alpha_s \nabla P - \nabla P_s - \beta(\vec{u_g} - \vec{u_s}) + \alpha_s \rho_s g \qquad (5)$$

(c) Granular Temperature

$$\Theta = \frac{1}{3}u'^2 \qquad (6)$$

(d) Equation of conservation of solids fluctuating energy

$$\frac{3}{2}\left(\frac{\partial(\alpha_s\rho_s\Theta)}{\partial t} + \nabla\cdot(\alpha_s\rho_s\vec{u}_s\Theta)\right) = \left(-P_s\bar{\bar{I}} + \bar{\bar{\tau}}_s\right):\nabla\vec{u}_s - \nabla\cdot q - \gamma - J \qquad (7)$$

(e) Equation of conservation of solids fluctuating energy in algebraic form

$$0 = \left(-P_s\bar{\bar{I}} + \bar{\bar{\tau}}_s\right):\nabla\vec{u}_s - \gamma_s \qquad (8)$$

**B. Constitutive equations**

(a) Gas phase stress tensor

$$\bar{\bar{\tau}}_g = \alpha_g\left[\left(\xi_g - \frac{2}{3}\mu_g\right)(\nabla\cdot\vec{u}_g)\bar{\bar{I}} + \mu_g\left((\nabla\vec{u}_g) + (\nabla\vec{u}_g)^T\right)\right] \qquad (9)$$

(b) Solid phase stress tensor

$$\bar{\bar{\tau}}_s = -\alpha_s\left[\left(\xi_s - \frac{2}{3}\mu_s\right)(\nabla\cdot\vec{u}_s)\bar{\bar{I}} + \mu_s\left((\nabla\vec{u}_s) + (\nabla\vec{u}_s)^T\right)\right] \qquad (10)$$

(c) Solid phase pressure

$$P_s = \alpha_s\rho_s\Theta + 2g_0\alpha_s^2\rho_s\Theta(1+e_s) \qquad (11)$$

(d) Solids shear viscosity

$$\mu_s = \mu_{s,col} + \mu_{s,kin} + \mu_{s,fr} \qquad (12)$$

(e) Collisional viscosity [13]

$$\mu_{s,col} = \frac{4}{5}\alpha_s\rho_s d_s g_0(1+e_s)\sqrt{\frac{\Theta}{\pi}} \qquad (13)$$

(f) Kinetic viscosity [13]

$$\mu_{s,kin} = \frac{10}{96}\sqrt{\Theta\pi}\frac{\rho_s d_s}{(1+e_s)\alpha_s g_0}\left[1+\frac{4}{5}g_0\alpha_s(1+e_s)\right]^2 \qquad (14)$$

(g) Kinetic viscosity [20]

$$\mu_{s,kin} = \frac{\alpha_s\rho_s d_s\sqrt{\Theta\pi}}{6(3-e_s)}\left[1+\frac{2}{5}g_0\alpha_s(1+e_s)(3e_s-1)\right] \qquad (15)$$

(h) Frictional viscosity [21]

$$\mu_{s,fr} = \frac{P_s\sin\phi}{2\sqrt{I_{2D}}} \qquad (16)$$

(i) Solids bulk viscosity [22]

$$\xi_s = \frac{4}{3}\alpha_s\rho_s d_s g_0 (1+e_s)\sqrt{\frac{\Theta}{\pi}} \qquad (17)$$

(j) Radial distribution function [22]

$$g_0 = \left[1-\left(\frac{\alpha_s}{\alpha_{s,max}}\right)^{1/3}\right]^{-1} \qquad (18)$$

(k) Collisional energy dissipation [22]

$$\gamma_s = 3(1-e_s{}^2)\alpha_s^2\rho_s g_0 \Theta \left(\frac{4}{d_p}\sqrt{\frac{\Theta}{\pi}}\right) \qquad (19)$$

(l) Gas–solid phase interphase exchange coefficient: Gidaspow drag model [13]

$\alpha_g > 0.8$
$$\beta = \frac{3}{4}C_{D0}\frac{\alpha_g(1-\alpha_g)}{d_s}\rho_g|\vec{u}_g - \vec{u}_s|\alpha^{-2.65} \qquad (20)$$
$\alpha_g \leq 0.8$

$$\beta = 150\frac{(1-\alpha_g)^2}{\alpha_g}\frac{\mu_g}{(d_s)^2} - 1.75(1-\alpha_g)\frac{\rho_g}{d_s}|\vec{u}_g - \vec{u}_s| \qquad (21)$$

$$C_{D0} = \begin{cases} \frac{24}{Re_s}[1+0.15(Re_s)^{0.687}], Re_s < 1000 \\ 0.44, Re_s > 1000 \end{cases} \qquad (22)$$

$$Re_s = \frac{\alpha_g\rho_g|\vec{u}_g - \vec{u}_s|d_s}{\mu_g} \qquad (23)$$

### 2.3. Initial and boundary conditions

For both the gas and solid particles' entries, the velocity inlet boundary condition was used while for the outlet, the pressure outlet boundary condition was applied. The system pressure was entered as an atmospheric pressure. For the solid particles, the initial solid inventory in the riser was set using available experimental data [10]. At the wall, the boundary condition of Johnson and Jackson [23] was applied which assumes no slip for the gas and partial slip for the solid phase. This condition has been successfully applied to KTGF modeling by Sinclair and Jackson [24]. Physical properties and simulation parameters are listed in Table 2.

*Table 2*. Modeling parameters

| Description | Value |
|---|---|
| Bed height H | 2 m |
| Bed width | 0.2 m |
| Static bed height $H_0$ | 0.1 m |
| Gas density $\rho_g$ | 1.2 Kg/m$^3$ |
| Particle density ρs | 2400 Kg/m$^3$ |
| Particle diameter ds | 120 μm |
| Initial solid volume fraction $\varepsilon_0$ | 0.6 |
| Inlet gas velocity $U_g$ | 1 m/s |
| Solid flux $G_s$ | 0.22 kg/m$^2$s |
| Angle of internal friction | 30° |
| Restitution coefficient $e_s$ | 0.9 |
| Specularity coefficient φ | 1 |
| Maximum particle packing limit | 0.64 |
| Time step | 10$^{-4}$s |

3. Results and discussion

In the current study, the system hydrodynamics were described in terms of the distribution of the solid volume fraction and the radial and axial velocities of the solid particles. The contours of the instantaneous solid concentration in the circulating fluidized bed at t=30 s for the non-modified and modified risers are given in Fig. 2. Two coexisting regions, characteristic of the turbulent fluidized bed (a bottom dense region and a top dilute region), are observed for all tested cases. The addition of ring baffles into the circulating fluidized bed has a significant effect on the distribution of the two phases. It makes, the solid volume fraction in both modified CFB risers evidently more homogeneous along the CFB riser column then in the reference case (case 1). In both modified risers, cases, most of the solid particles flew upwards at the center region and only few solid particles fell downwards near the wall region. This is due to the presence of the ring baffles that improves the breaking of the clusters near the wall region and pushes the solid particles towards the center of the flow. Plots on Fig.3 to Fig.8 are time-averaged results at the quasi-state condition, over 10 to 30 seconds of simulation time. For instance, Fig.3 present the time-averaged solid volume fractions along the height of CFB riser under different ring baffle configurations. The Time-averaged solid volume fraction for the two modified configurations is much lower than that of the reference configuration, thus justifying the significant effect of the geometrical structure on the particles' flow. Fig. 3 also shows a uniform time-averaged solid volume fraction over a longer distance (1.2-2 m) in the riser with comparison to the reference configuration (1.6-2 m), indicating a better axial dispersion of the solids. We can also observe from these plots that the cross-sectional mean solid holdup for case 2 and case 3 are confused with the existence of a moderate difference in the values reached. This observation suggests that the shape of the ring baffles has a limited impact on the distribution of the particles in the center of the column.

In addition to the solid volume fraction, the pressure drop is a further parameter that requires full attention when adding ring baffles into the riser since the flow interruption from the ring baffles may induce the system pressure to drop or the energy requirement to increase. The computed time-averaged system pressure drops along the height of non-modified and modified CFB risers are shown in Fig. 4. In absence of ring baffles, the system pressure drop profile for the case without ring baffles has a smooth exponential shape which divided into two distinct zones: the acceleration zone and the established zone. At the bottom of the bed, the entering particles are accelerated by gas flow. Due to a relatively higher solids' holdup and the particle acceleration, this region displays a high-system pressure drop. This part of the riser is called the acceleration zone. Beyond this acceleration region, the system pressure drop does not significantly change. This implies that at the upper part of the riser both the particle velocity and the solid holdup become constant, this is the established zone. The presence of the baffles in the fluidization column affects significantly the axial distribution of the pressure drop by the introduction of pressure drop peaks at the location of the ring baffles and tends to decrease at other locations which is mainly due to the high particle retention and the particles' accelerations effects around the rings, as explained by Bu and Zhu [25]. This explanation assumes that the particles accumulated in the baffle region absorb more kinetic energy to climb into the column, which results in an increased pressure drop in this area. The solids acceleration zone is inherent to the circulation fluidized beds. However, it is undesirable for many reactions because of the increased back-mixing in this relatively dense region. The addition of ring baffles seems to sensibly reduce the length of this acceleration section. In fact, this action considerably reduced the length of this zone from 1.4 m for the reference case (case 1) to 1 m for the two configurations with baffles. The installation of the baffles improves then the quality of axial mixing within the fluidization column, probably due to the narrower path way created by the ring baffles, which leads to a higher gas velocity and then assists to the acceleration of solids. On the other hand, it should be noted that the implementation of ring baffles is pronounced only at the acceleration section, whereas, in the top developed section, the pressure appears to be similar for all profiles. We finally note that the shape of the ring baffles has a limited impact on the distribution of the system pressure drop (similar plots and close peak values).

The system mixing can also be idescribed by means of the axial and radial solid particle and gas velocity distributions. The axial distributions of the time-averaged axial solid particle and gas velocities of the CFB riser with different ring baffle configuration are displayed in Fig. 5 where positive value means the upward movement while the negative values mean a downward movement. The introduction of ring baffles has a significant effect, expressed through a higher level of fluctuation along the axial distribution of the different plotted velocities. In fact, when the solid particles pass through the baffle zone, they change direction and promoting fluctuations along the axial velocity profiles. The flow of the two phases becomes consequently much more turbulent, contributing then to the increase of the intensity of the axial mixing.

The axial distribution of the time-averaged radial velocities of both solid particles gas within the CFB riser under different ring baffle configurations are plotted in Fig. 6, where positive values represent a left to right particle movement and negative values a right to left movement. For all three riser geometries, the radial gas and solid particle velocities are lower than the axial ones, since the axial direction is the main direction of the system flow. The addition of the ring baffles increases the solid particles' movement, through higher positive and negative radial solid particle and gas velocities. In fact, when the solid particles reach the ring baffles positions, they are induced into the center region, a radial movement resulting in a sign change in both radial velocities along the CFB riser height. The effect of the ring baffles introduction over the radial velocities is further emphasized in the bottom of the riser. In fact, for the two modified configurations, we can clearly observe the high positive and negative radial solid particle and gas velocities. These variations are due to the combination of an increase in the gas velocity in the riser center due to the ring baffles and the movement of solid particle from the inlet to the opposite wall. Due to this secondary solids flow, the particles circulate though the walls and are dragged from one side to the other. Therefore, cases 2 and 3 had a low opportunity to form particle clusters near the wall region in the bottom zone of the riser, promoting as a result the radial gas-solid particle mixing. As we rise in the column, the fluctuation decrease sensibly and the profiles converge towards the basic configuration profile with small discrepancies in the values reached. This suggests that the improvement of the radial mixture by the installation of baffles is more pronounced in the lower zone of the column characterized by a strong retention of solid particles.

The radial distribution of the time–averaged axial solid particles' velocity within the CFB riser with and without ring baffles at two axial positions between the rings (H=0.5 m and H=1.3 m) are illustrated in Fig.7. The overall observation of these plots reveals a central region characterized by an upward movement of the solid particles (positive velocities) and an annular region characterized by a downward movement for all configurations. Cases 2 and 3 o have the lowest negative axial solid particles' velocities due to having the lowest level of solid particles falling downwards, near the wall region. The ring baffles block consequently the solid particles fall down and force them to move to the center region that likely leads to a lower solid back mixing than the reference case (case 1). We can also see from these plots that for the cases 2 and 3, the thickness of the central region (0.01<X<0.19) is extended reflecting a better quality of the mixing in this region. We can also note in this figure that for the two modified configurations, the profiles of the time-averaged axial solid particle velocity are almost coincide with slight difference in the values reached. This suggests that the shape of the baffles has limited impact on the radial distribution of the axial solid particles' velocity.

The radial distributions of the time-averaged radial solid particles' velocities in the CFB risers with and without ring baffles for two axial positions between the rings (H=0.5 m and H=1.3 m) are displayed in Fig.8. The plots highlight the preponderance of the axial flow, since the velocity values of the solid particles in the axial direction (Fig.7) are much greater than those in the radial direction. We can also see from these plots that the modification made on the CFB riser has a significant effect on the solid particles' flow in the radial direction. Indeed, the cases 2 and 3 have more fluctuating profiles than the case without ring baffles as indicated by the numerically larger positive and negative values of the radial solid particles' velocity at the right and left sides of the system. The increase of the radial velocity is more pronounced in the transition region between the core and the annular region that reached up to 0.4 m/s at H=0.5 m , which reflects the existence of a privileged area for the transfer of mass from the core to the annular region. Indeed, the brusque enlargement of cross-sectional area of the riser caused by the shape of the baffles contributes to the formation of recirculation zones at the corners and therefore to the entrainment of the two phases towards the walls. This flow pattern induces a significant increase of the velocity in the transition zone between the core and the annular and thus the improvement of the mixing quality in the radial direction. This statement is comforted by Fig.9, where the streamlines of the gas phase for the two modified configurations reveal the formation of the vortex (recirculation zone) structures developed just above the baffles. From Fig.8, we can also observe that the shape of the baffles affects the radial distribution of the radial velocity. In fact, for case 3, the mean radial solid particle velocities are higher than those of the case 2. This suggests that the case 3 has a low possibility to form clusters and therefore a great possibility, to improve the gas-solids mixture quality in the radial direction.

## 4. Conclusion

An Eulerian-Eulerian CFD model incorporating the Kinetic Theory of Granular Flow (KTGF) was applied by means of the commercial CFD package Ansys Fluent in order to explore the effect of ring baffle configuration on the fluidization system hydrodynamics in a CFB riser filled with Geldart B particles. In this approach, two different types of ring baffles, circular and trapezoidal, were considered. As can be seen from the simulation results, the insertion of ring baffles into a CFB riser has a significant effect on the flow patterns. The ring baffles improved the system mixing in both radial and axial direction and reduced the backflow near the wall, which allowed a higher benefit for heat and mass transfer in this CFB riser system. It was also found that the shape of the ring baffles had a limited impact on the system hydrodynamics.

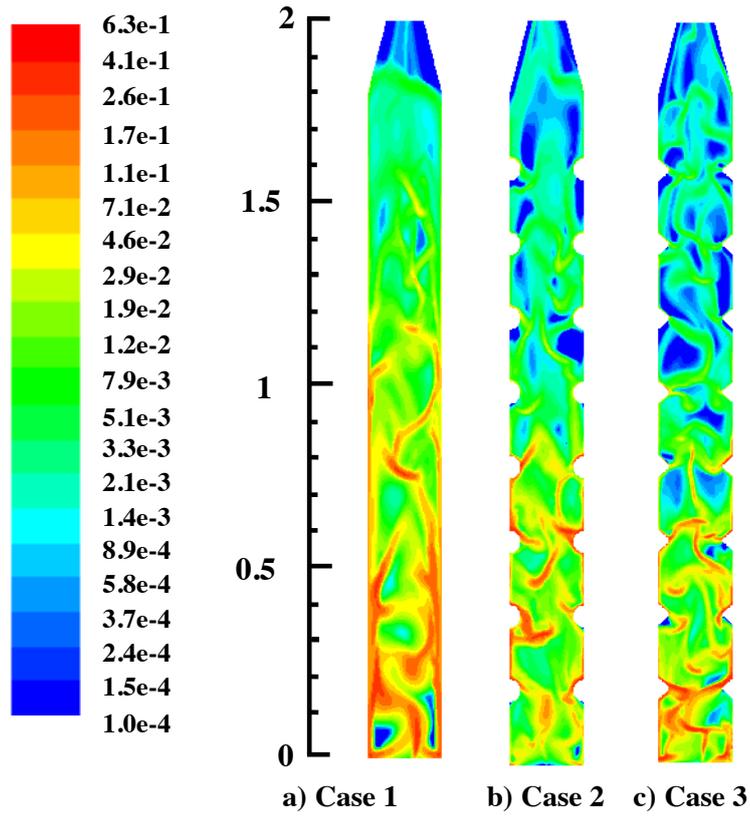

***Fig. 2.*** *The contours of instantaneous solid volume fraction at 30 s in CFB riser: (a) without baffles (b) with circular ring baffles and (c) with trapezoidal ring baffles*

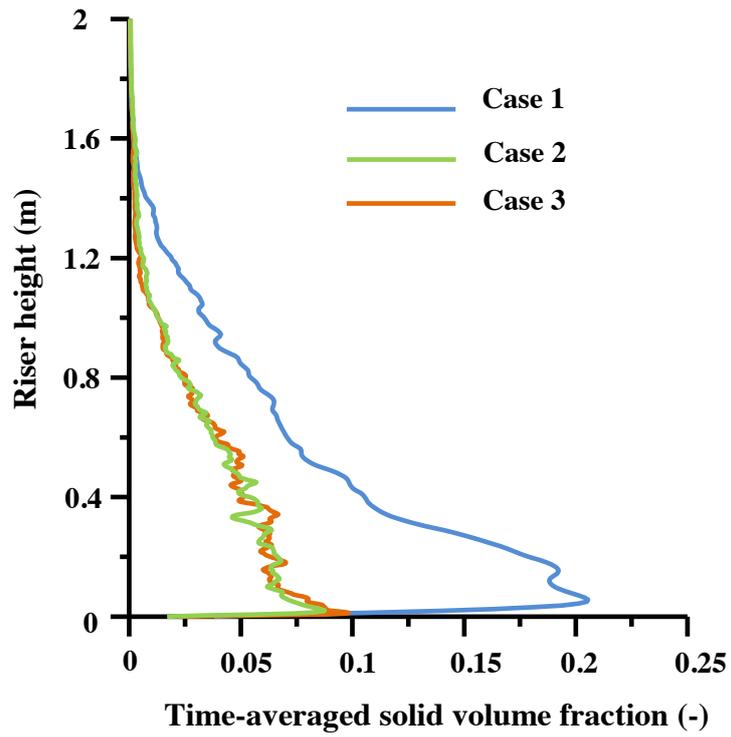

*Fig. 3.* *The time-averaged solid volume fractions along the height of CFB riser: without baffles (case 1), with circular ring baffles (case 2) and with trapezoidal ring baffles (case 3)*

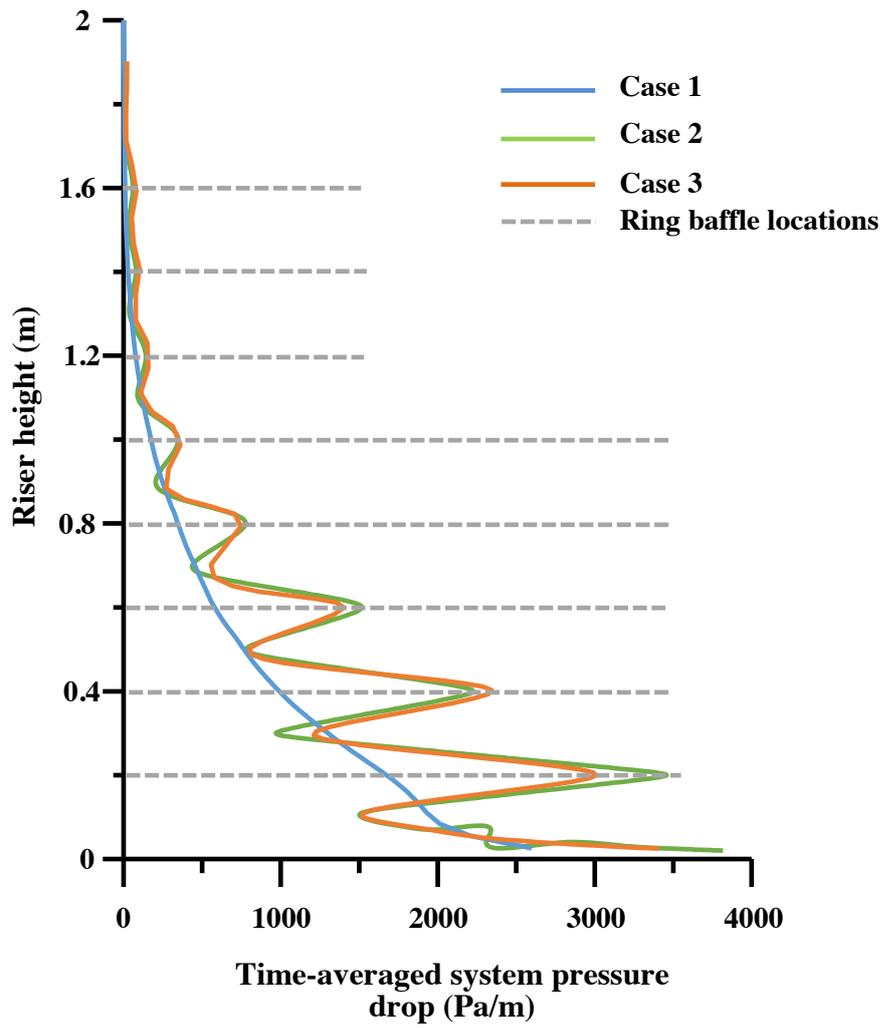

*Fig. 4.* *The time-averaged system pressures drop along the height of CFB riser: without baffles (case 1), with circular ring baffles (case 2) and with trapezoidal ring baffles (case 3)*

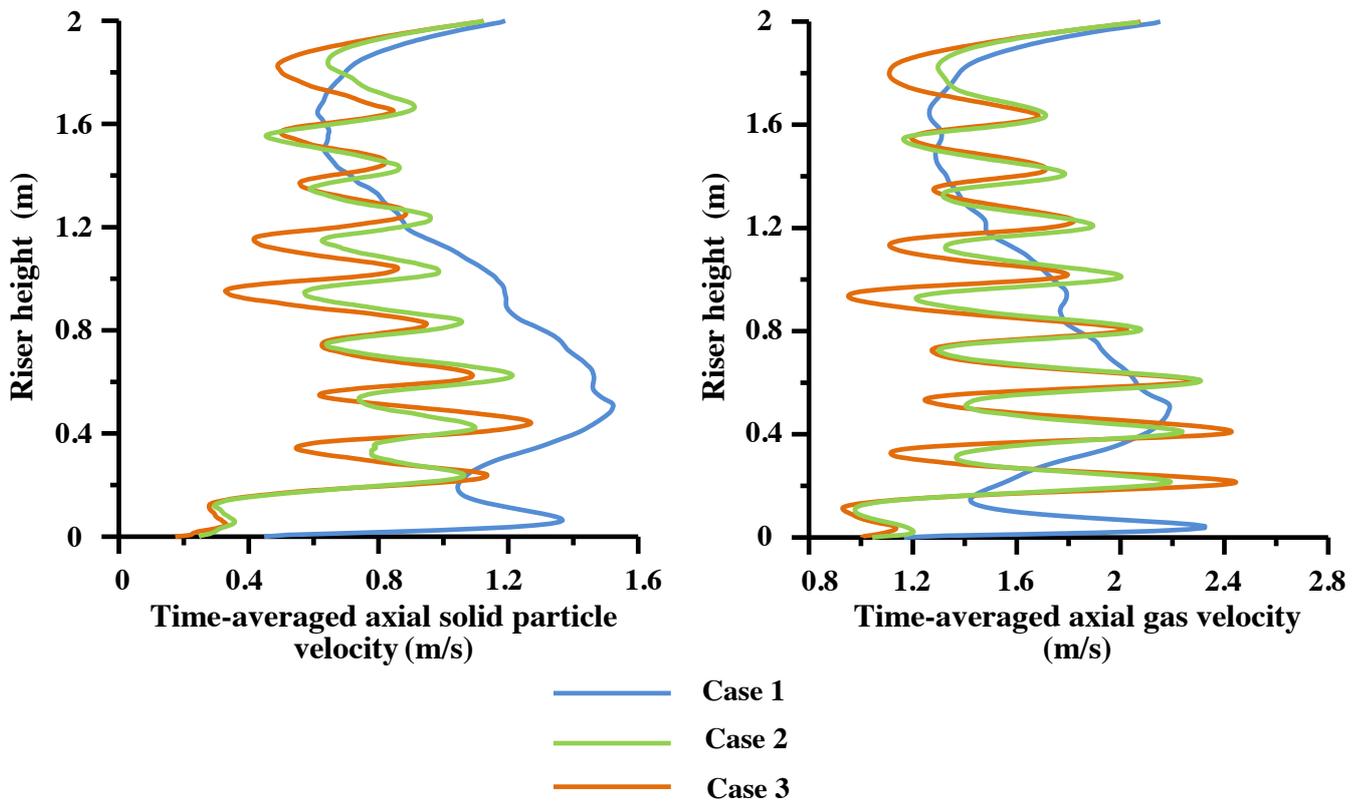

*Fig. 5.* *The axial distributions of time-averaged axial velocities of : a) solid particles and b) gas along the height of CFB riser with different ring baffle configurations.*

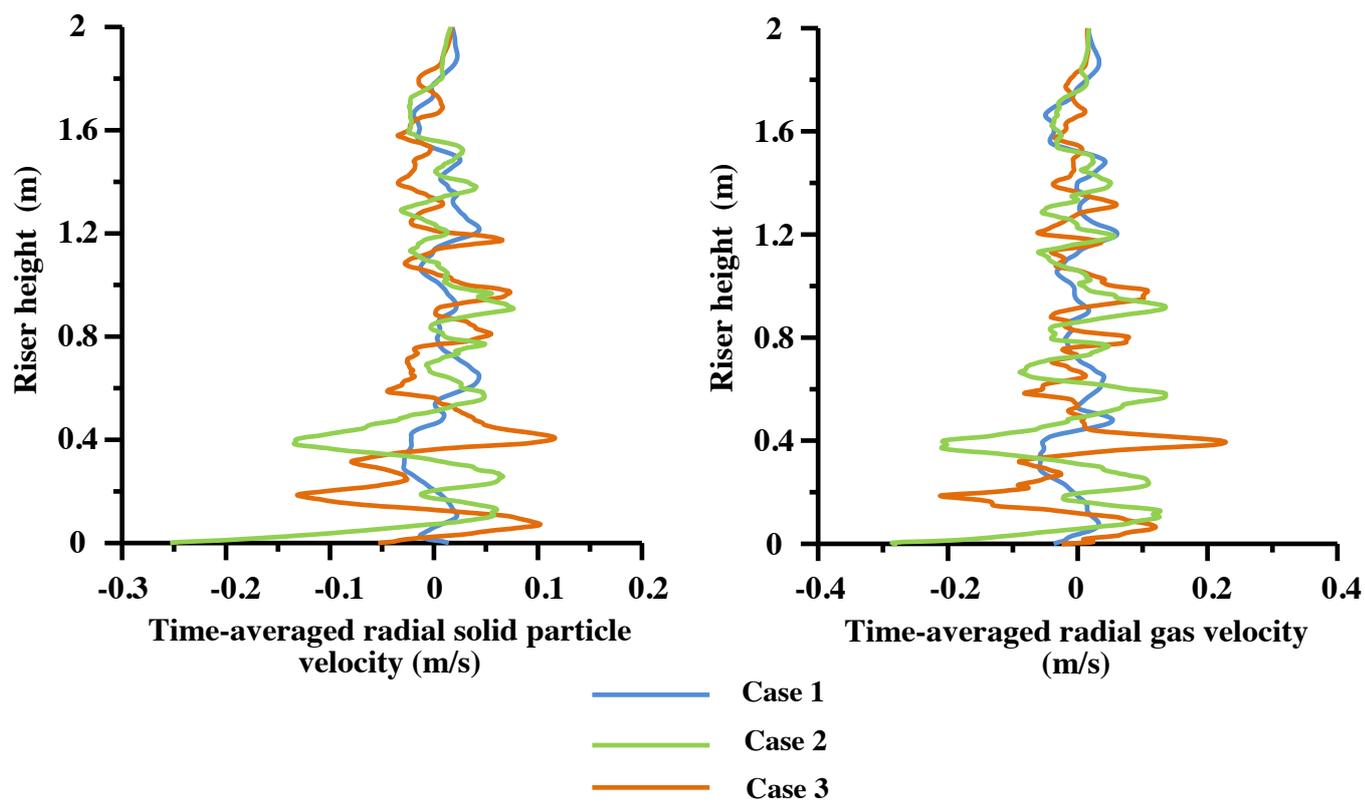

*Fig. 6.* The axial distributions of time-averaged radial velocities of : a) solid particles and b) gas along the height of CFB riser with different ring baffle configurations.

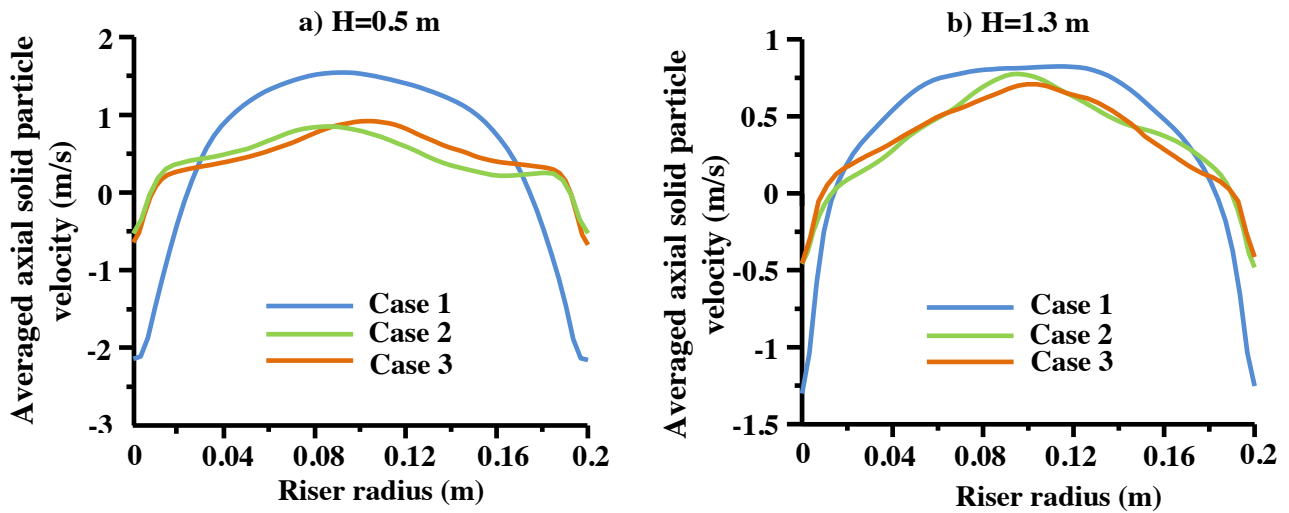

***Fig. 7.*** *The radial distributions of time-averaged axial solid particle velocity at two axial positions of the CFB riser with different ring baffle configurations : (a) H=0.5 m and (b) H=1.3 m*

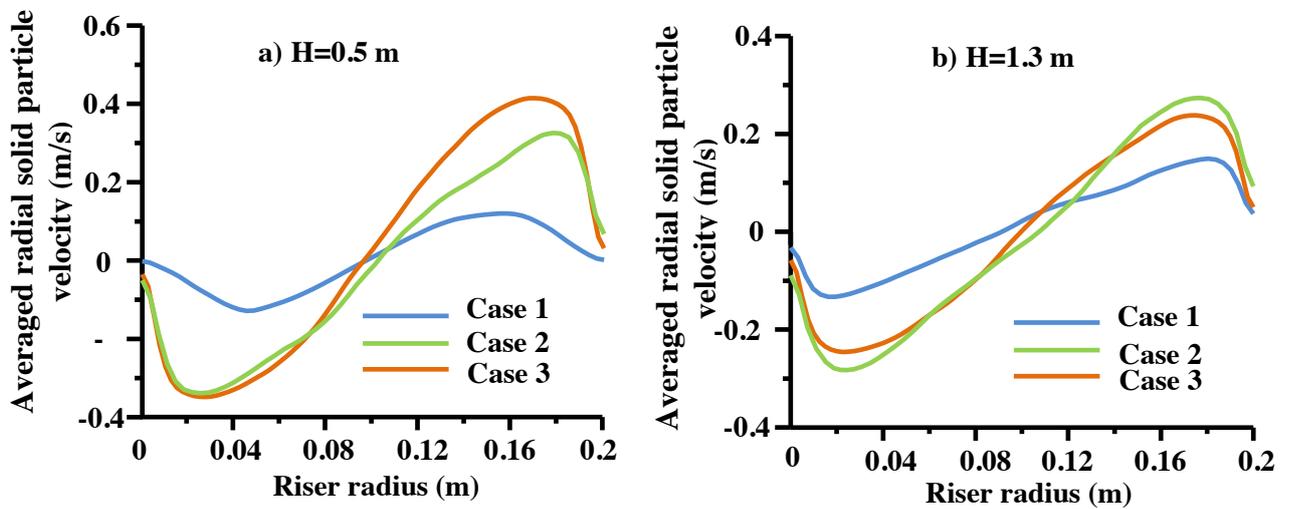

***Fig. 8.*** *The radial distributions of time-averaged radial solid particle velocity for two axial positions of the CFB riser with different ring baffle configurations : (a) H=0.5 m and (b) H=1.3 m.*

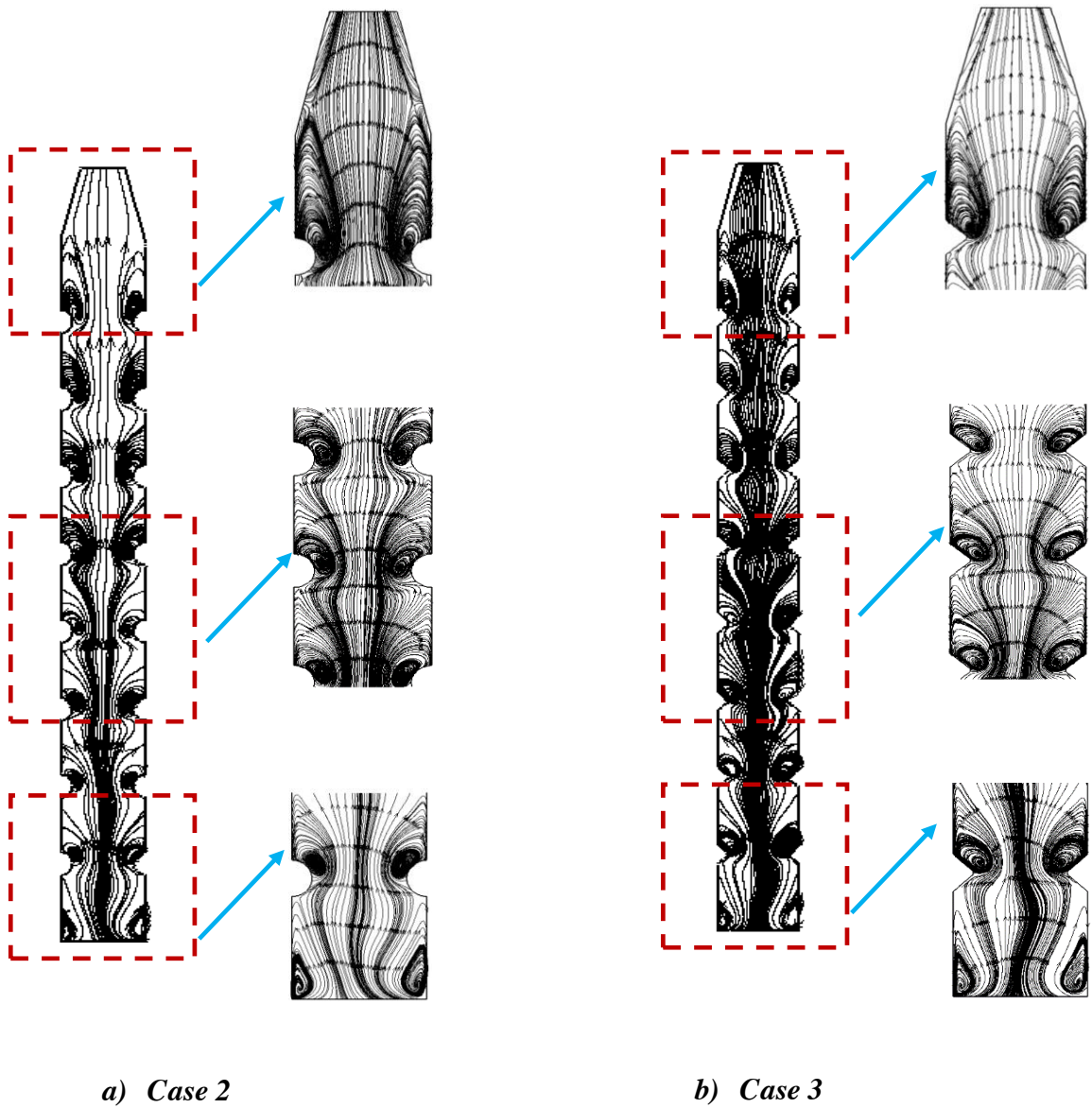

*Fig. 9.* Streamlines of gas phase for the two modified configurations: (a) with circular ring baffles and (b) with trapezoidal ring baffle.

**References**
[1] J.R. Grace, A.A. Avidan, T.M. Knowlton, Circulating Fluidized Beds, Blackie Academic and Professional, London, UK (1997).
[2] W. Namkung, S.D. Kim, Gas backmixing in a circulating fluidized bed, Powder Technol. 99 (1998) 70-78.
[3] M.V.D. Velden, J. Baeyens, K. Smolders, Solids mixing in the riser of a circulating fluidized bed, Chem. Eng. Sci. 62 (2007) 2139-2153.
[4] P. Jiang, H. Bi, R. Jean, L. Fan, Baffle effect on performance of catalytic circulating fluidized bed reactor, AIChE J. 37 (1991) 1392-1400.
[5] C. Zheng, Y. Tung, Y. Xia, B. Hun, M. Kwauk, Voidage redistribution by ring internals in fast fluidization. In: Dwauk, M., Hasatani, M. (Eds.), Fluidization'91, Science and Technology, Science Press, Beijing, (1991) 168-177.
[6] J. Zhu, M. Salah, Y. Zhou, Radial and axial voidage distributions in circulating fluidized bed with ring-type internals, J. Chem. Eng. Jpn. 30 (1997) 928-937.
[7] A. Therdthianwong, P. Pantaraks, S. Therdthianwong, Modeling and simulation of circulating fluidized bed reactor with catalytic ozone decomposition reaction, Powder Technol. 133 (2003) 1-14.
[8] T. Samruamphianskun, P. Piumsomboon, B. Chalermsinsuwan, Effect of ring baffle configurations in a circulating fluidized bed riser using CFD simulation and experimental design analysis, Chem. Eng. J. 210 (2012) 237-251.
[9] T. Samruamphianskun, P. Piumsomboon, Effect of operating parameters inside circulating fluidized bed reactor riser with ring baffles using CFD simulation and experimental design analysis, Chemical Engineering Research and Design. 92 (2014) 2479-2492.
[10] A. Zaabout, H. Bournot, R. Occelli, B. Kharbouch, Local solid particle behavior inside the upper zone of a circulating fluidized bed riser, Adv. Powder Technol. 22 (2011) 375–382.
[11] D. Kunii, O. Levenspiel, Fluidization Engineering, Butterworth-Heinemann, Newton, United States (1991).
[12] S. Benzarti, H. Mhiri, H. Bournot, R. Occelli, Numerical simulation of turbulent fluidized bed with Geldart B particles, Adv. Powder Technol. 25 (2014) 1737-1747.
[13] D. Gidaspow, Multiphase Flow and Fluidization: Continuum and Kinetic Theory Descriptions, Academic Press, San Diego (1994).
[14] B. Sun, D. Gidaspow, Computation of circulating fluidized-bed riser flow for thefluidization VIII benchmark test, Ind. Eng. Chem. Res. 38 (1999) 787-792.
[15] Neri, A., Gidaspow, 2000. Riser hydrodynamics: simulation using kinetic theory, AIChE J. 46, 52-67.
[16] L. Huilin, D. Gidaspow, Hydrodynamic simulations of gas–solid flow in a riser, Ind. Eng. Chem. Res. 42 (2003) 2390-2398.
[17] Z. Yunhau, L. Huilin, H. Yurong, J. Ding, Y. Lijie, Numerical prediction of combustion of carbon particle clusters in a circulating fluidized bed riser, Chem. Eng. J. 118 (2006) 1-10.
[18] R. Andreux, G. Petit, M. Hemati, O. Simonin, Hydrodynamic and solid residence time distribution in a circulating fluidized bed: Experimental and 3D computational study, Chem. Eng. Process. 47 (2007) 463-473.
[19] J.O. Hinze, Turbulence, McGraw-Hill, New York (1975).
[20] M. Syamlal, T.J. O'Brien, Computer simulation of bubbles in a fluidized bed, AIChE Symp. Ser. 85 (1989) 22-31.

[21] D.G. Schaeffer, Instability in the evolution equations describing incompres- sible granular flow, Journal of differential equations 66 (1987) 19-50.

[22] C. Lun, S. Savage, D. Jeffrey, N. Chepumiy, Kinetic theories of granular flow: inelastic particles in coquette flow and slightly inelastic particles in a general flow field, Journal of Fluid Mechanics. 140 (1984) 223-256.

[23] P.C. Johnson, P. Nott, R. Jackson, Frictional–collisional equations of motion for participate flows and their application to chutes, J.Fluid Mech. 210 (1990) 501-535.

[24] J.L. Sinclair, R. Jackson, Gas–particle flow in a vertical pipe with particle–particle interaction, AIChE J. 35 (1989) 1473-1486.

[25] J. Bu, J. X. Zhu, Influence of ring-type internals on axial pressure distribution in circulating fluidized bed, Can. J. Chem. Eng. 77 (1999), 26-34.

**Fig. 1.** *Schema of the different CFBR tested geometries: (a) without baffles (b) with circular ring baffles and (c) with trapezoidal ring baffles*

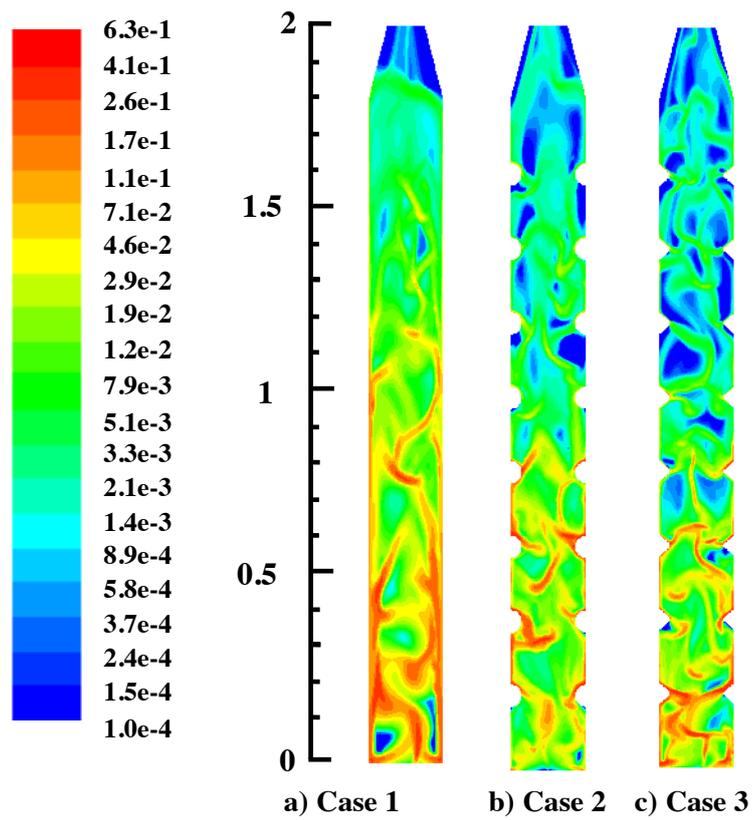

***Fig. 2.*** *The contours of instantaneous solid volume fraction at 30 s in CFB riser: (a) without baffles*

*(b) with circular ring baffles and (c) with trapezoidal ring baffles*

**Figure 3**

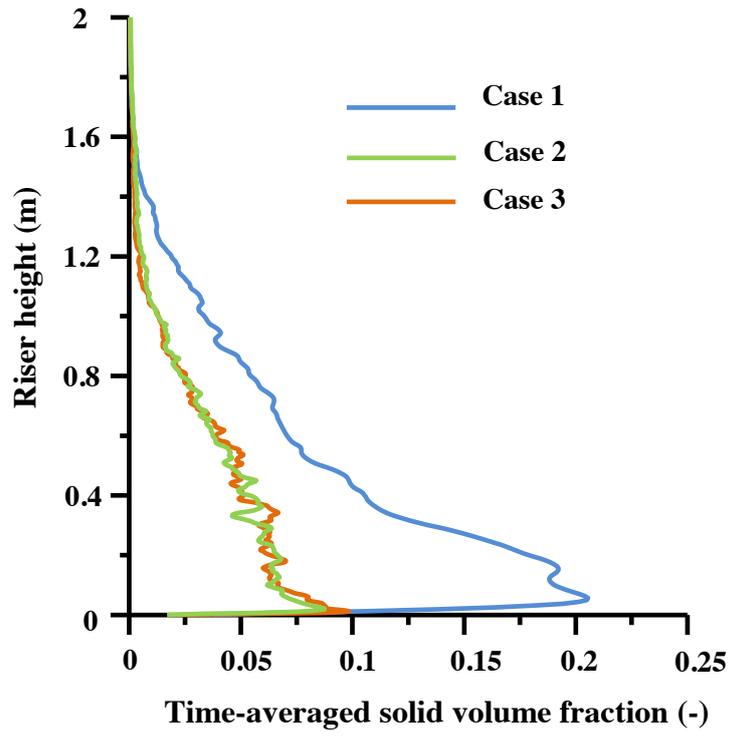

*Fig. 3.* The time-averaged solid volume fractions along the height of CFB riser: without baffles (case 1), with circular ring baffles (case 2) and with trapezoidal ring baffles (case 3)

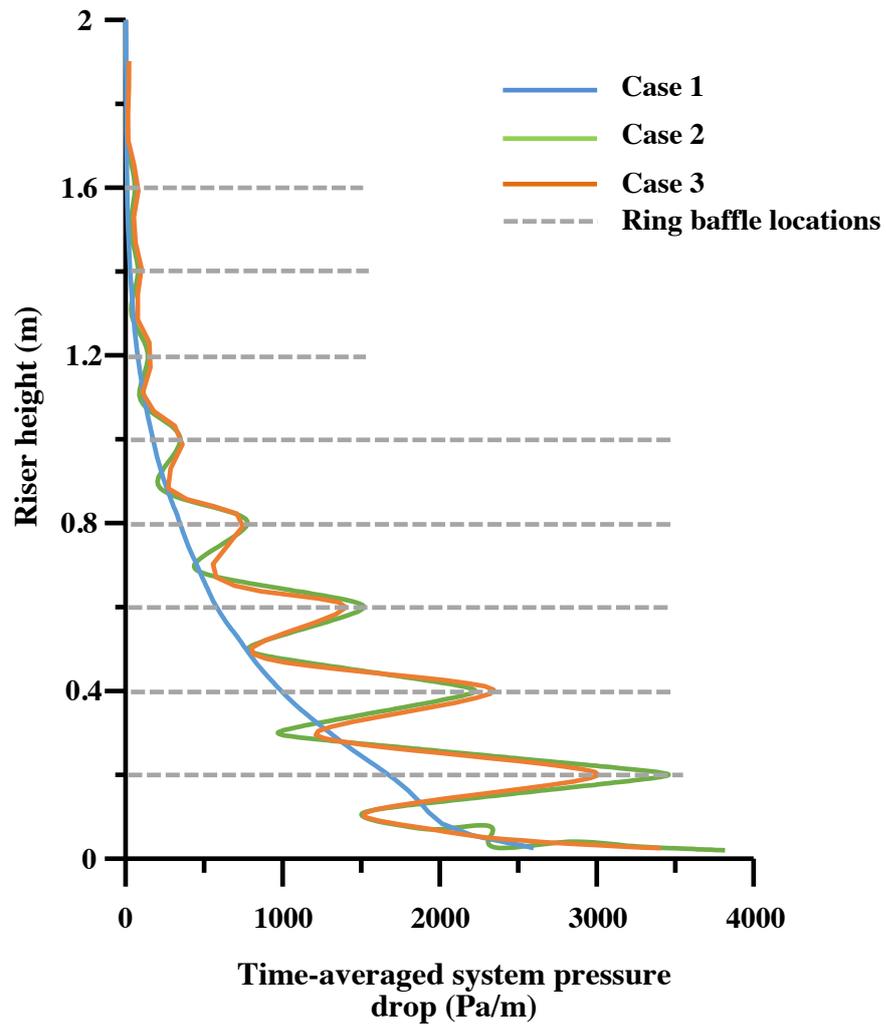

*Fig. 4.* The time-averaged system pressures drop along the height of CFB riser: without baffles (case 1), with circular ring baffles (case 2) and with trapezoidal ring baffles (case 3)

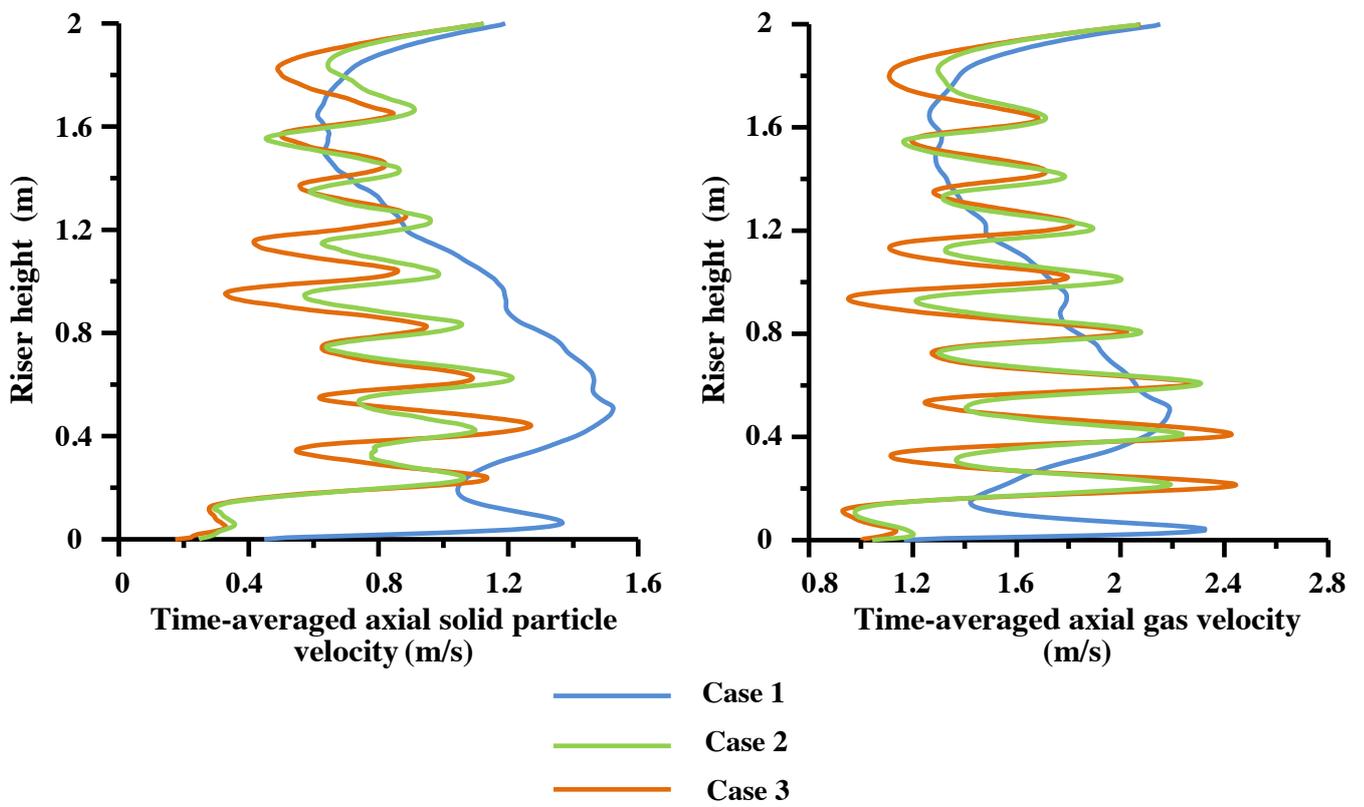

*Fig. 5.* *The axial distributions of time-averaged axial velocities of : a) solid particles and b) gas along the height of CFB riser with different ring baffle configurations.*

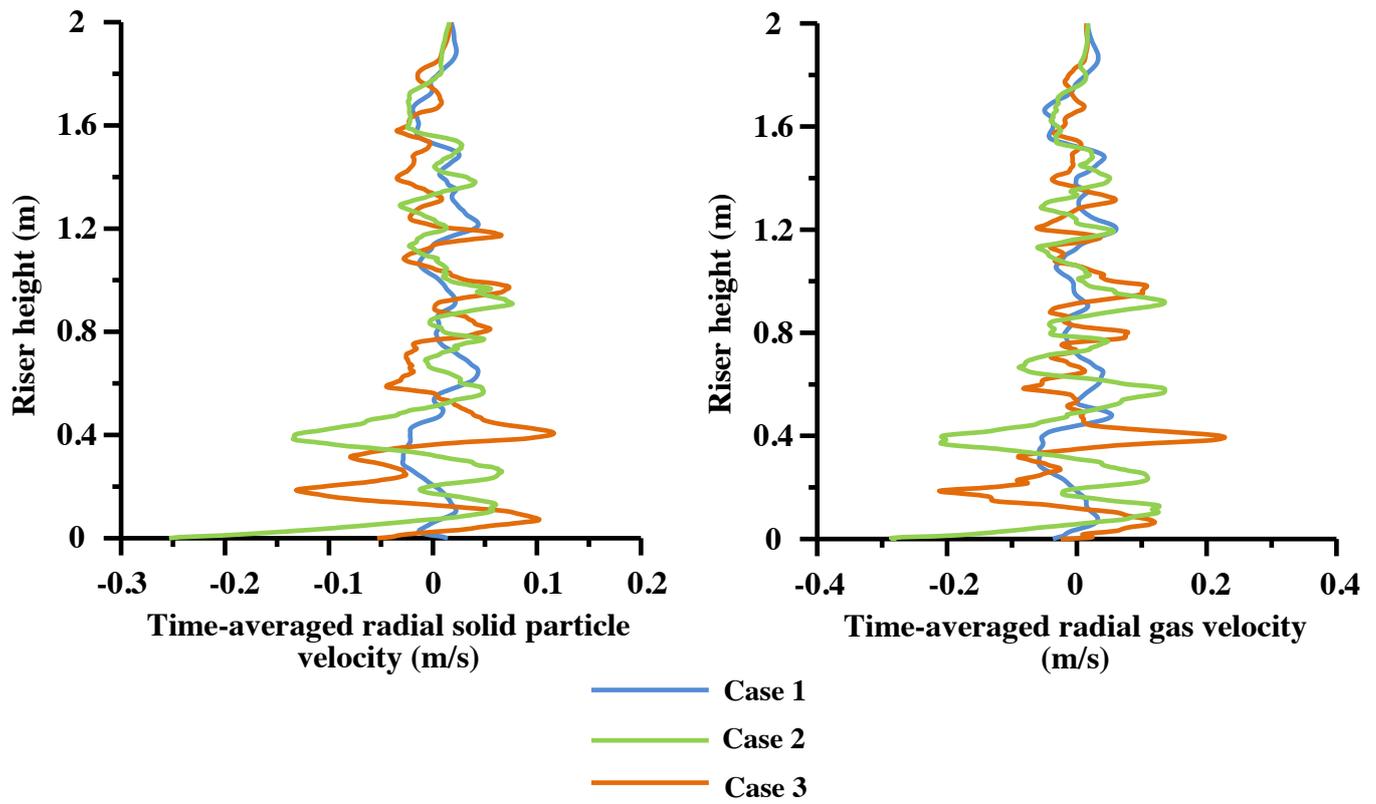

*Fig. 6.* *The axial distributions of time-averaged radial velocities of : a) solid particles and b) gas along the height of CFB riser with different ring baffle configurations.*

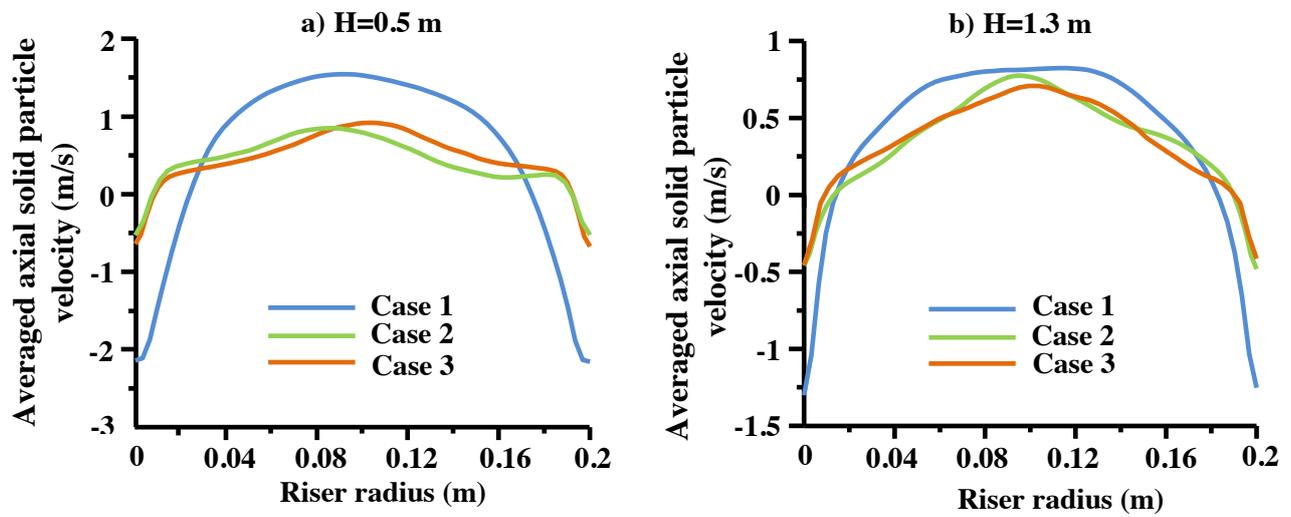

*Fig. 7.* *The radial distributions of time-averaged axial solid particle velocity at two axial positions of the CFB riser with different ring baffle configurations : (a) H=0.5 m and (b) H=1.3 m*

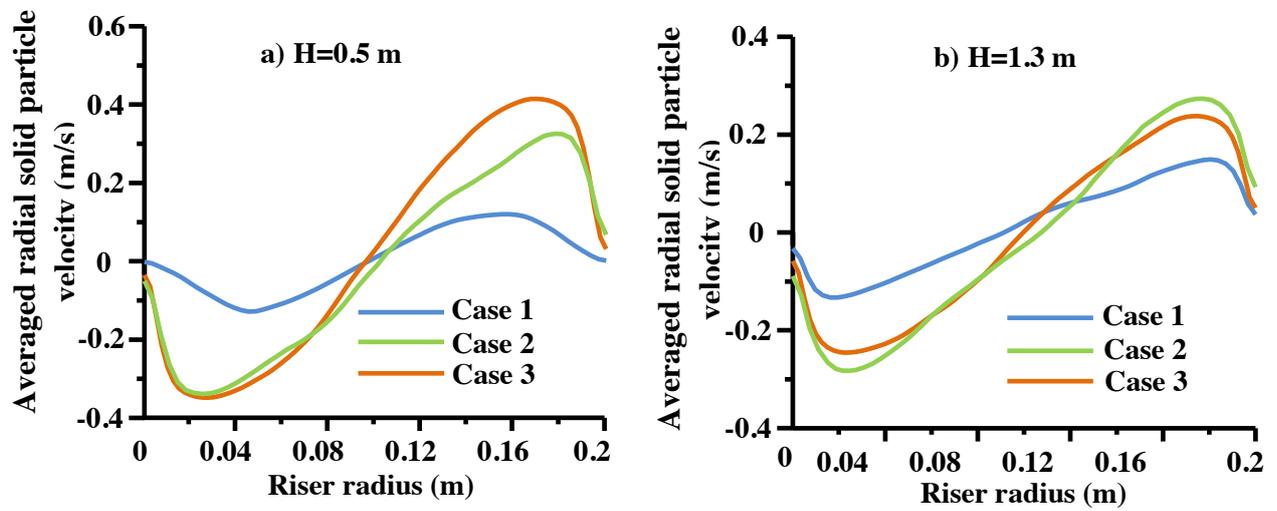

*Fig. 8.* *The radial distributions of time-averaged radial solid particle velocity for two axial positions of the CFB riser with different ring baffle configurations : (a) H=0.5 m and (b) H=1.3 m.*

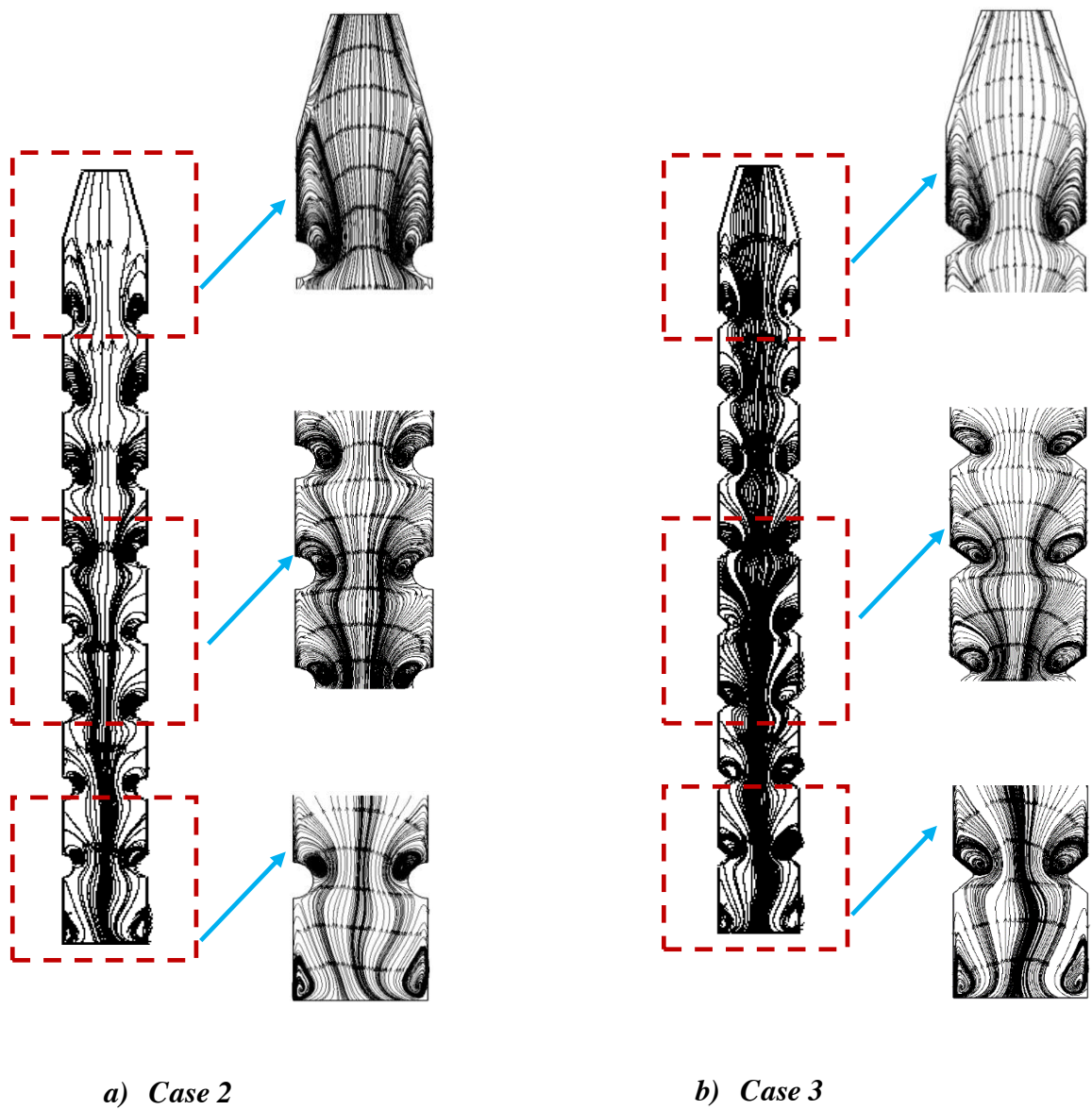

*a) Case 2*  *b) Case 3*

***Fig. 9.*** *Streamlines of gas phase for the two modified configurations: (a) with circular ring baffles and (b) with trapezoidal ring baffle.*

**Table 1**
A summary of the governing equations and constitutive equations

### A. Governing equations

(a) Mass conservation equations of gas and solids phases

$$\frac{\partial(\alpha_g \rho_g)}{\partial t} + \nabla \cdot (\alpha_g \rho_g \vec{u}_g) = 0 \tag{1}$$

$$\frac{\partial(\alpha_s \rho_s)}{\partial t} + \nabla \cdot (\alpha_s \rho_s \vec{u}_s) = 0 \tag{2}$$

$$\alpha_g + \alpha_s = 1 \tag{3}$$

(b) Momentum conservation equations of gas and solids phases

$$\frac{\partial(\alpha_g \rho_g \vec{u}_g)}{\partial t} + \nabla \cdot (\alpha_g \rho_g \vec{u}_g \vec{u}_g) = \nabla \cdot (\bar{\bar{\tau}}_g) - \alpha_g \nabla P - \beta(\vec{u}_g - \vec{u}_s) + \alpha_g \rho_g g \tag{4}$$

$$\frac{\partial(\alpha_s \rho_s \vec{u}_s)}{\partial t} + \nabla \cdot (\alpha_s \rho_s \vec{u}_s \vec{u}_s) = \nabla \cdot (\bar{\bar{\tau}}_s) - \alpha_s \nabla P - \nabla P_s - \beta(\vec{u}_g - \vec{u}_s) + \alpha_s \rho_s g \tag{5}$$

(c) Granular Temperature

$$\Theta = \frac{1}{3} u'^2 \tag{6}$$

(d) Equation of conservation of solids fluctuating energy

$$\frac{3}{2}\left(\frac{\partial(\alpha_s \rho_s \Theta)}{\partial t} + \nabla \cdot (\alpha_s \rho_s \vec{u}_s \Theta)\right) = (-P_s \bar{\bar{I}} + \bar{\bar{\tau}}_s) : \nabla \vec{u}_s - \nabla \cdot q - \gamma - J \tag{7}$$

(e) Equation of conservation of solids fluctuating energy in algebraic form

$$0 = (-P_s \bar{\bar{I}} + \bar{\bar{\tau}}_s) : \nabla \vec{u}_s - \gamma_s \tag{8}$$

### B. Constitutive equations

(a) Gas phase stress tensor

$$\bar{\bar{\tau}}_g = \alpha_g \left[\left(\xi_g - \frac{2}{3}\mu_g\right)(\nabla \cdot \vec{u}_g)\bar{\bar{I}} + \mu_g\left((\nabla \vec{u}_g) + (\nabla \vec{u}_g)^T\right)\right] \tag{9}$$

(b) Solid phase stress tensor

$$\bar{\bar{\tau}}_s = -\alpha_s \left[\left(\xi_s - \frac{2}{3}\mu_s\right)(\nabla \cdot \vec{u}_s)\bar{\bar{I}} + \mu_s\left((\nabla \vec{u}_s) + (\nabla \vec{u}_s)^T\right)\right] \tag{10}$$

(c) Solid phase pressure

$$P_s = \alpha_s \rho_s \Theta + 2g_0 \alpha_s^2 \rho_s \Theta(1 + e_s) \tag{11}$$

(d) Solids shear viscosity

$$\mu_s = \mu_{s,col} + \mu_{s,kin} + \mu_{s,fr} \tag{12}$$

(e) Collisional viscosity [13]

$$\mu_{s,col} = \frac{4}{5}\alpha_s\rho_s d_s g_0(1+e_s)\sqrt{\frac{\Theta}{\pi}} \quad (13)$$

(f) Kinetic viscosity [13]

$$\mu_{s,kin} = \frac{10}{96}\sqrt{\Theta\pi}\frac{\rho_s d_s}{(1+e_s)\alpha_s g_0}\left[1+\frac{4}{5}g_0\alpha_s(1+e_s)\right]^2 \quad (14)$$

(g) Kinetic viscosity [20]

$$\mu_{s,kin} = \frac{\alpha_s\rho_s d_s\sqrt{\Theta\pi}}{6(3-e_s)}\left[1+\frac{2}{5}g_0\alpha_s(1+e_s)(3e_s-1)\right] \quad (15)$$

(h) Frictional viscosity [21]

$$\mu_{s,fr} = \frac{P_s\sin\phi}{2\sqrt{I_{2D}}} \quad (16)$$

(i) Solids bulk viscosity [22]

$$\xi_s = \frac{4}{3}\alpha_s\rho_s d_s g_0(1+e_s)\sqrt{\frac{\Theta}{\pi}} \quad (17)$$

(j) Radial distribution function [22]

$$g_0 = \left[1-\left(\frac{\alpha_s}{\alpha_{s,max}}\right)^{1/3}\right]^{-1} \quad (18)$$

(k) Collisional energy dissipation [22]

$$\gamma_s = 3(1-e_s^2)\alpha_s^2\rho_s g_0\Theta\left(\frac{4}{d_p}\sqrt{\frac{\Theta}{\pi}}\right) \quad (19)$$

(l) Gas–solid phase interphase exchange coefficient: Gidaspow drag model [13]

$\alpha_g > 0.8$

$$\beta = \frac{3}{4}C_{D0}\frac{\alpha_g(1-\alpha_g)}{d_s}\rho_g|\vec{u}_g-\vec{u}_s|\alpha^{-2.65} \quad (20)$$

$\alpha_g \leq 0.8$

$$\beta = 150\frac{(1-\alpha_g)^2}{\alpha_g}\frac{\mu_g}{(d_s)^2} - 1.75(1-\alpha_g)\frac{\rho_g}{d_s}|\vec{u}_g-\vec{u}_s| \quad (21)$$

$$C_{D0} = \begin{cases} \frac{24}{Re_s}[1+0.15(Re_s)^{0.687}], Re_s < 1000 \\ 0.44, Re_s > 1000 \end{cases} \quad (22)$$

$$Re_s = \frac{\alpha_g\rho_g|\vec{u}_g-\vec{u}_s|d_s}{\mu_g} \quad (23)$$

**Table 2**

*Table 2. Modeling parameters*

| Description | Value |
|---|---|
| Bed height H | 2 m |
| Bed width | 0.2 m |
| Static bed height $H_0$ | 0.1 m |
| Gas density $\rho_g$ | 1.2 Kg/m$^3$ |
| Particle density ρs | 2400 Kg/m$^3$ |
| Particle diameter ds | 120 μm |
| Initial solid volume fraction $\varepsilon_0$ | 0.6 |
| Inlet gas velocity $U_g$ | 1 m/s |
| Solid flux $G_s$ | 0.22 kg/m$^2$s |
| Angle of internal friction | 30° |
| Restitution coefficient $e_s$ | 0.9 |
| Specularity coefficient φ | 1 |
| Maximum particle packing limit | 0.64 |
| Time step | 10$^{-4}$s |

*Highlights*

- The circulating fluidized bed with ring baffles was investigated.
- Eulerian–Eulerian approach with Gidaspow gas-solid drag model and standard k-e model was performed
- Two ring baffle configurations were investigated.
- Time-averaged results relative to circulating fluidized beds in presence and in absence of ring baffles were compared.
- The incorporation of ring baffles improved the system mixing and reduced the backflow near the wall

**\*Declaration of Interest Statement**

**Declaration of interests**

☒ The authors declare that they have no known competing financial interests or personal relationships that could have appeared to influence the work reported in this paper.

☐The authors declare the following financial interests/personal relationships which may be considered as potential competing interests: